\documentclass[aps,prl,showpacs,superscriptaddress,amssymb,longbibliography,twocolumn]{revtex4-1}
\usepackage{graphicx}
\usepackage{appendix} 
\usepackage{bm,amsmath}
\usepackage{dcolumn}
\usepackage{hyperref}

\renewcommand{\vec}[1]{\boldsymbol{#1}}  

\usepackage{color}
\newcommand{\zo}[1]{{\color[RGB]{0, 0, 0}{#1}}}
\newcommand{\ats}[1]{{\color[RGB]{0, 0, 0}{#1}}}

\begin{document}

\rightline{\tt }

\vspace{0.2in}

\title{Experimental Quantum Learning of a Spectral Decomposition}

\author{Michael R. Geller}
\thanks{mgeller@uga.edu}
\affiliation{Center for Simulational Physics, University of Georgia, Athens, Georgia 30602, USA}

\author{Zo\"{e} Holmes} 
\affiliation{Information Sciences, Los Alamos National Laboratory, Los Alamos, NM, USA.}

\author{Patrick J. Coles}
\affiliation{Theoretical Division, Los Alamos National Laboratory, Los Alamos, NM, USA.}
\affiliation{Quantum Science Center, Oak Ridge, TN, 37931, USA.}

\author{Andrew Sornborger} 
\thanks{sornborg@lanl.gov}
\affiliation{Information Sciences, Los Alamos National Laboratory, Los Alamos, NM, USA.}
\affiliation{Quantum Science Center, Oak Ridge, TN, 37931, USA.}


\begin{abstract}
Currently available quantum hardware allows for small scale implementations of quantum machine learning algorithms. Such experiments aid the search for applications of quantum computers by benchmarking the near-term feasibility of candidate algorithms. Here we demonstrate the quantum learning of a two-qubit unitary by a sequence of three parameterized quantum circuits containing a total of 21 variational parameters. Moreover, we variationally diagonalize the unitary to learn its spectral decomposition, i.e., its eigenvalues and eigenvectors. We illustrate how this can be used as a subroutine to compress the depth of dynamical quantum simulations. One can view our implementation as a demonstration of entanglement-enhanced machine learning, as only a single (entangled) training data pair is required to learn a $4\times 4$ unitary matrix.
\end{abstract}

\maketitle

Quantum simulation and machine learning are among the most promising applications of large-scale quantum computers. Of these, the discovery of algorithms with provable exponential speedup has been more challenging in the machine learning domain, in part because it is harder to port established machine learning techniques to the quantum setting \cite{biamonte2017quantum}. Notable exceptions include linear system solvers \cite{harrow2009quantum,clader2013preconditioned,childs2017quantum}, kernel methods \cite{schuld2019quantum,havlivcek2019supervised}, and Boltzmann machines \cite{wiebe2014quantum,verdon2017quantum}. But quantum simulation demonstrations \cite{kandala2017hardware,kandala2019error,1084,arute2020observation,aleiner2020accurately} appear to be ahead of machine learning \cite{riste2017demonstration,schuld2017implementing,yao2017quantum,huang2018demonstration,Zhueaaw9918,kathuria2020implementation,tomesh2020coreset} in terms of the maximum problem sizes achieved, suggesting that quantum simulation might yield the earliest applications with quantum advantage. 

Variational quantum algorithms~\cite{cerezo2020variationalreview,endo2021hybrid,bharti2021noisy} will likely facilitate near-term implementations for these applications. Such algorithms employ a problem-specific cost function
that is evaluated on a quantum computer, while a
classical optimizer trains a parameterized quantum circuit to minimize this cost.

Some variational quantum algorithms have interest beyond their ability to achieve quantum advantage on their own, and can serve as subroutines for larger quantum algorithms. These include quantum autoencoders for data compression~\cite{romero2017quantum,bondarenko2020quantum} and linear algebra methods~\cite{bravo2019variational,xu2019variational,huang2019near}. A common subroutine is the training of a variational quantum state to approximate the ground state of a given $n$-qubit Hamiltonian, which can be the Hamiltonian of a simulated model \cite{peruzzo2014variational} or some other optimization objective \cite{farhi2014quantum,hadfield2019quantum,mitarai2018quantum}. Variational quantum algorithms to learn and diagonalize density matrices have also been developed \cite{larose2019variational,cerezo2020variational,bravo2020quantum}, which is a fundamental subroutine that will have many uses including principal component analysis and estimation of quantum information quantities~\cite{cerezo2020variationalfidelity,beckey2020variational}. Another subroutine is the learning of one quantum state by a second state, where the output of a variational circuit is optimized to maximize the overlap with an input state that might itself be the output of another algorithm \cite{khatri2019quantum,sharma2020noise,jones2018quantum,cincio2018learning}. Although a minimum of $2(2^n-1)$ real parameters are required to do this exactly in general, it is widely believed that for many cases of interest a polynomial number of parameters will suffice.

Beyond learning states, it is also useful to \textit{variationally learn a unitary channel} $\rho \mapsto U\rho U^\dagger$. This is more challenging because now \ats{all of the $4^n-1$ matrix elements must match} for an exact replica $V$ of an arbitrary $U$, up to a global phase. A hybrid protocol for learning a unitary is provided by the quantum-assisted quantum compiling algorithm \cite{khatri2019quantum}. This is a low-depth subroutine appropriate for both near-term and fault-tolerant quantum computing.

\zo{Given a target unitary $U$ and parameterized unitary $V(\theta)$, both acting on $n$-qubits, quantum-assisted quantum compiling uses a maximally entangled Bell state on $2n$ qubits to compute the Hilbert-Schmidt inner product, $| {\rm Tr}(UV(\theta)^\dagger)|$. Since this inner product is directly related to the average fidelity between states acted on by $U$ and $V(\theta)$~\cite{HHH99, nielsen02}, this allows the action of $U$ on all possible input states to be studied using a single entangled input state. Consequently, with this entanglement-enhanced learning strategy, only a single training state is needed to fully learn $U$, in contrast to the ${\footnotesize \sim}2^{n}$ input-output pairs that are required in the absence of entanglement \cite{poland2020no,sharma2020reformulation}.}

Although more challenging than state learning, learning a unitary can be used for a wide variety of quantum information applications, including circuit depth compression, noise-tailoring,  benchmarking, and the `black box uploading' of an unknown experimental system unitary \cite{khatri2019quantum}. Quantum-assisted quantum compiling has been demonstrated on 1+1 qubits, where a single-qubit variational circuit $V(\theta)$ learned the value of a single-qubit unitary $U$ \cite{khatri2019quantum}.

Quantum-assisted quantum compiling can be generalized to learn not only a unitary, but also its Schur decomposition $W(\theta) D(\gamma) W(\theta)^\dagger$, where $W(\theta)$ is a parameterized unitary and $D(\gamma)$ is a parameterized diagonal unitary. \zo{That is, one can use quantum-assisted quantum compiling to \textit{variationally diagonalize a unitary}. This is useful for a variety of quantum information science applications, since access to the spectral decomposition of a unitary $U$ enables arbitrary powers of $U$ to be implemented using a fixed depth circuit. Specifically, suppose we learn the optimum parameters $\theta_{\rm opt}$ and $\gamma_{\rm opt}$ such that $U = W(\theta_{\rm opt}) D(\gamma_{\rm opt}) W(\theta_{\rm opt})^\dagger$. We can then implement $U^k$ using the fixed depth circuit $W(\theta_{\rm opt}) D(k \gamma_{ \rm opt}) W(\theta_{\rm opt})^\dagger$. We stress that the parameter $k$ here can take any real value and hence this approach can be used to implement non-integer and negative powers of $U$. }

\zo{One important application of variational unitary diagonalization is quantum simulation.} Let $U$ be a Trotterized (or other) approximation to a short-time evolution $e^{-i H \Delta t}$ by some Hamiltonian $H$. We assume that $H$ is local~\cite{lloyd1996universal,childs2019nearly,haah2021quantum,low2019well}, sparse~\cite{aharonov2003adiabatic,berry2014exponential,low2017optimal}, or given by a linear combination of unitaries~\cite{berry2015simulating,berry2015hamiltonian,low2019hamiltonian}, permitting efficient simulation with bounded error. Then $W D^{t/ \Delta t} W^\dagger$ is an approximate {\it fast-forwarded} evolution operator with a circuit depth that is independent of $t$.
By contrast, most of the best known Hamiltonian simulation algorithms~\cite{low2019well,low2017optimal,low2019hamiltonian} have depths scaling at least linearly in $t$, inhibiting long time simulations on near-term hardware. 

This low-depth algorithm, called variational fast-forwarding~\cite{cirstoiu2020variational}, lies at an exciting intersection of machine learning and quantum simulation and is an promising approach in the burgeoning field of variational quantum simulation~\cite{arrasmith2019variational,trout2018simulating,endo2020variational,yao2020adaptive, benedetti2020hardware, heya2019subspace, bharti2020quantum,lau2021quantum,haug2020generalized, barison2021efficient,gibbs2021long,commeau2020variational}. 
Variational fast-forwarding has been demonstrated on 1+1 qubits~\cite{cirstoiu2020variational}. Refinements of variational fast-forwarding for simulating a given fixed initial state~\cite{gibbs2021long} and for learning the spectral decomposition of a given Hamiltonian~\cite{commeau2020variational} have also been proposed.
\zo{It is important to note that the unitary being diagonalized need not already be known. Therefore, variational unitary diagonalization could also be used to perform a `black box diagonalization' of the dynamics of an unknown experimental system. Thus, this approach provides a new algorithmic tool for probing dynamics in an experimental setting.}


In this work we use ibmq\_bogota to demonstrate the successful learning of a spectral decomposition on 2+2 qubits. \zo{Specifically, we diagonalize the short time evolution unitary for an Ising spin chain.} After only 16 steps of training by gradient descent, the spectral decomposition is used to fast-forward the evolution of the Ising model, resulting in a dramatic reduction of simulation error compared with Trotterized Hamiltonian simulation and a significant ($\sim \! 10 \times$) increase in the effective quantum volume of the simulation.

\section{Methods}

\begin{figure*}
\includegraphics[width=11.0cm]{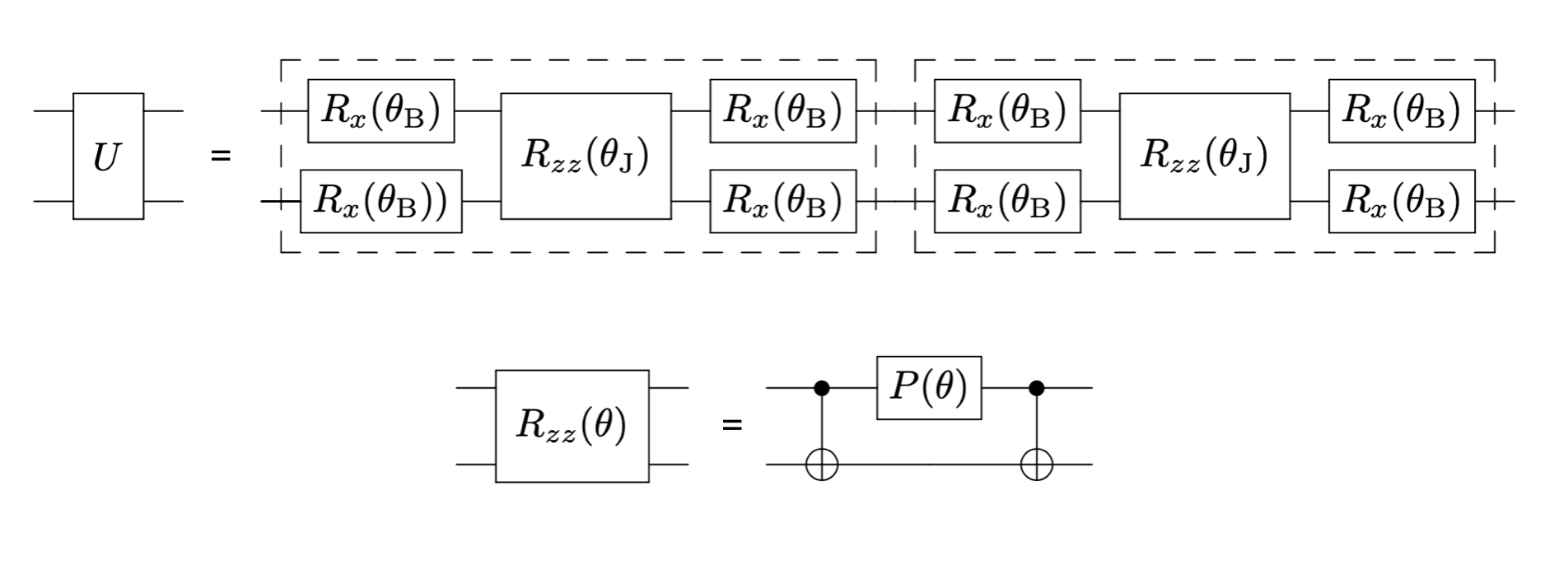} 
\caption{\textbf{Circuit for the short-time evolution of the Ising model.} The boxes indicate individual Trotter steps. Here $R_{\rm x}(\theta) = e^{i \theta X/2}$ is a single qubit Pauli $X$ rotation, $P(\gamma)={\rm diag}(1,e^{i \gamma})$ is a single qubit phase gate and $R_{\rm zz}(\theta) = e^{i \theta Z \otimes Z/2}$ is a two-qubit diagonal unitary.}
\label{U circuit figure}
\end{figure*}

\paragraph*{Learning task.}
We demonstrate the variational learning of the spectral decomposition of a unitary by learning a diagonalization of the short-time evolution operator of the $2$-spin Ising model
\begin{equation}
H =  J  \sum_{i=1,2} Z_{i} Z_{i+1} + B  \sum_{i=1,2}  X_i \, .
\label{defIsing}
\end{equation}
Here $J$ quantifies the exhange energy, $B$ is the transverse field strength and $Z_i$ and $X_i$ are Pauli operators on the $i_{\rm th}$ qubit.
We approximate the short-time evolution $\exp(- i H \Delta t)$ of the spin chain using a second order Trotter approximation, that is we take
\begin{equation}
U =  \bigg[ \prod_i \! R_{\rm x}(\theta_{\rm B})_i
 \times \prod_i \! R_{\rm zz}(\theta_{\rm J})_{i,i+1}  \times
\prod_i \! R_{\rm x}(\theta_{\rm B})_i   \bigg]^2 
\label{shot time trotter}
\end{equation}
where $\theta_{\rm B} \!=\! B \Delta t / 2$ and $\theta_{\rm J} \!=\! 2 J  \Delta t / 2$. The simulated Ising model parameters are listed in Table~\ref{parameter table}. The specific circuit we used for $U$ is shown in Fig.~\ref{U circuit figure}. 

\begin{table}[htb]
\centering
\caption{Ising model parameters.}
\begin{tabular}{|c|c|c|}
\hline
 & {\bf Parameter} & {\bf Value} \\
\hline 
$N$ & number spins & 2 \\
\hline 
$\Delta t$ & short evolution time & 0.2 \\
\hline 
$J$ & exchange energy & 1 \\
\hline 
$B$ & transverse field & 1 \\
\hline 
\end{tabular}
\label{parameter table}
\end{table}

\medskip
\paragraph*{Ansatz.}

To learn the spectral decomposition of $U$ we variationally compile it to a unitary with a
structure of the form
\begin{equation}
    V(\vec{\theta}, \vec{\gamma}) = W(\vec{\theta}) D(\vec{\gamma}) W(\vec{\theta})^\dagger \, ,
\end{equation}
where $W$ is an arbitrary unitary, $D$ is a diagonal matrix and  $\vec{\theta}$ and $\vec{\gamma}$ are vectors of parameters. After successful training, $D$ will capture the eigenvalues of $U$, and $W$ will capture the rotation matrix from the computational basis to the eigenbasis of $U$.

\begin{figure*}
\includegraphics[width=15.0cm]{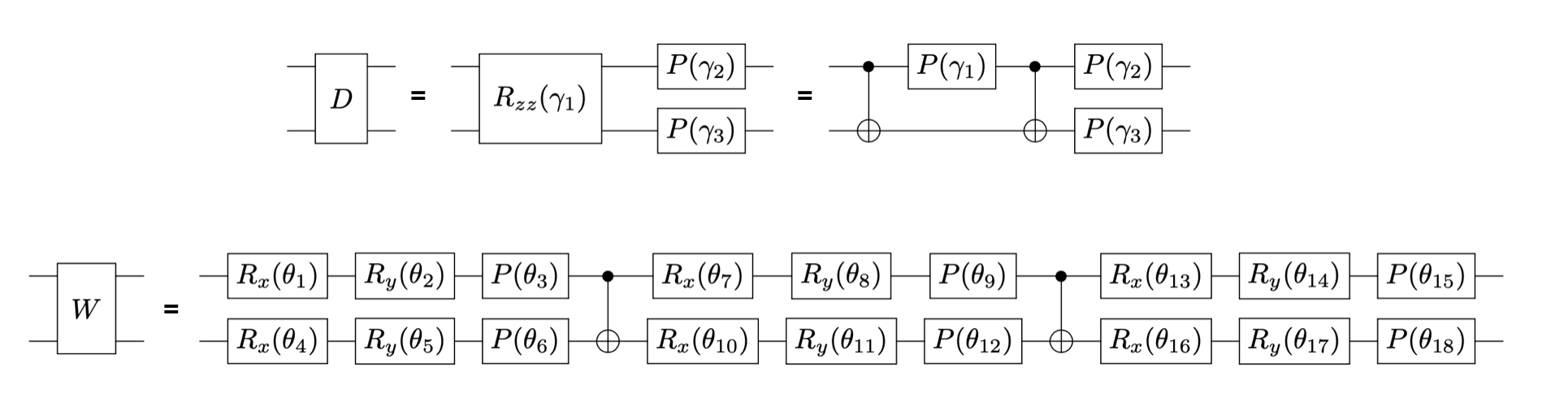} 
\caption{\textbf{Ansatz circuits.} Implemented circuits for the diagonal unitary $D$ and basis transformation $W$. Here $R_{\rm zz}(\gamma) = e^{i \gamma Z \otimes Z/2}$ is a two-qubit diagonal unitary and $P(\gamma)={\rm diag}(1,e^{i \gamma})$ is a single qubit phase gate.  $D$ contains 3 variational parameters. $W$ has 3 layers and 18 variational parameters.}
\label{DW circuit figure}
\end{figure*}

The parameterized circuits we used as ans\"{a}tze for the diagonal unitary $D$ and basis transformation $W$ are shown in Fig.~\ref{DW circuit figure}. A general diagonal unitary $D \in {\rm SU}(2^n)$ on $n$ qubits contains $2^n - 1$ variational parameters. In our experiment we implement a two-qubit version of this exact $D$ containing 3 variables.
In general an arbitrary unitary $W$ can be constructed from any general $n$-qubit parameterized quantum circuit with ${\rm poly}(n)$ variables. The expressive power of different options has been investigated in \cite{tangpanitanon2020expressibility,sim2019expressibility}. The two-qubit circuit we use to implement the arbitrary unitary $W$ consists of 3 layers of $X$, $Y$ rotations and phase gates on each qubit, separated by 2 CNOT gates.

\medskip

\begin{figure}
\includegraphics[width=8.5cm]{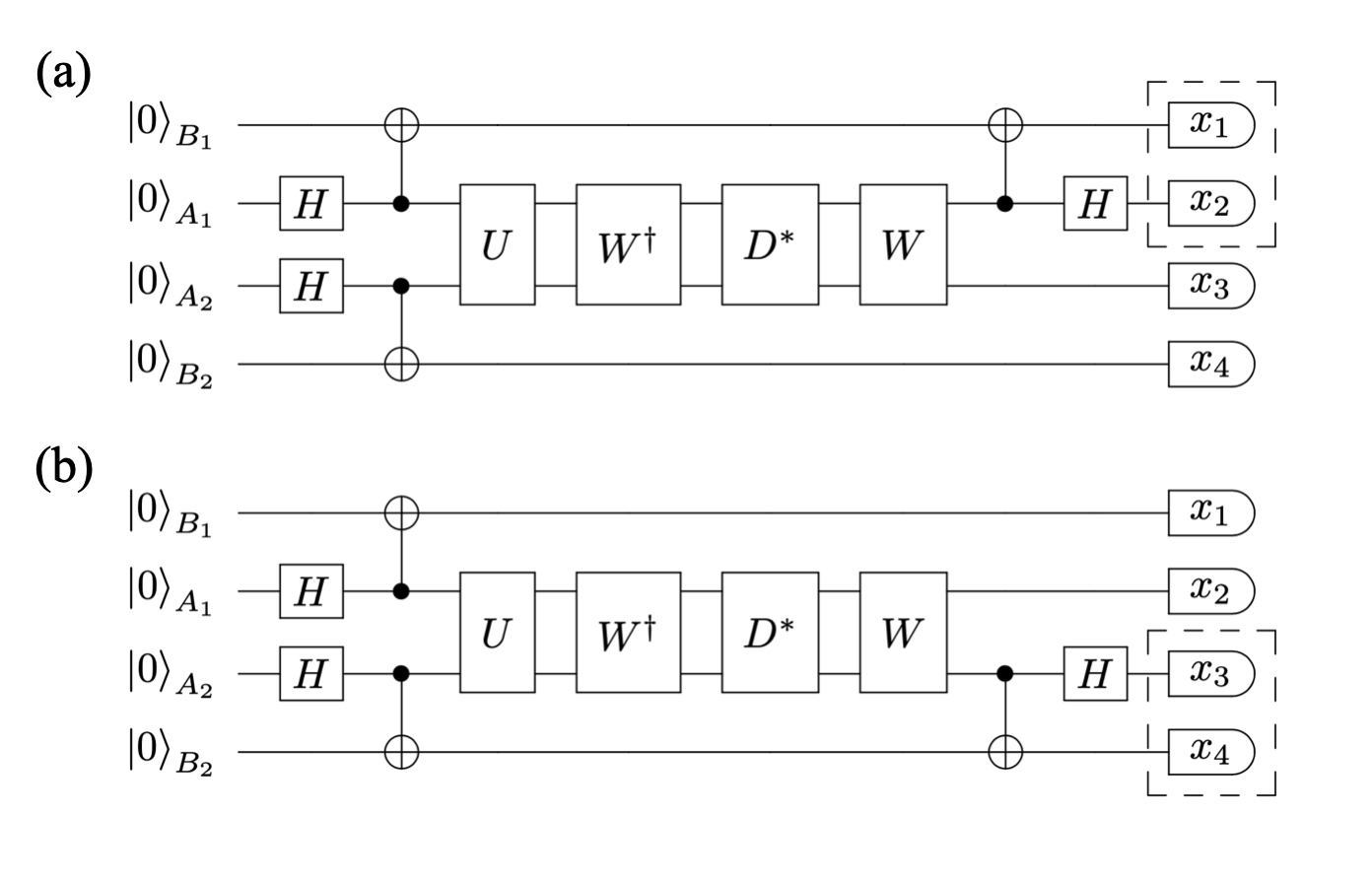}
\caption{\textbf{Cost function circuit.} Circuits to learn the spectral decomposition $WDW^\dagger$ of a two-qubit unitary $U$ on a four-qubit chain. Each cost function and gradient evaluation requires the estimation of  ${\rm Pr}(00)_{1,2}$ and ${\rm Pr}(00)_{3,4}$, obtained by measuring the qubits in the dashed boxes of (a) and (b).}
\label{spectral decomp figure}
\end{figure}

\paragraph*{Cost function.}
To compile the target unitary into a diagonal form we use the local Hilbert-Schmidt test cost function defined in \cite{khatri2019quantum}. For learning a $4 \times 4$ unitary matrix, this cost can be written as
\begin{equation}
C = 1 - \frac{1}{2} \left( {\rm Pr}(00)_{1,2} + {\rm Pr}(00)_{3,4}  \right) \, 
\label{defC}
\end{equation}
where $ {\rm Pr}(00)_{1,2}$ and $ {\rm Pr}(00)_{3,4}$ are the probabilities of observing the outcome $00$ on qubits (1,2) and (3,4) on running the circuits shown in Fig.~\ref{spectral decomp figure}(a) and Fig.~\ref{spectral decomp figure}(b) respectively. 

The probabilities ${\rm Pr}(00)_{1,2}$ and ${\rm Pr}(00)_{3,4}$ are measures of the entanglement fidelity of the unitary channel $U V^\dagger$ with $V = W D W^\dagger$. As a result, this cost is faithful, vanishing if and only if the diagonalization $W D W^\dagger$ matches the target unitary $U$ (up to a global phase). Furthermore, the cost is operationally meaningful for non-zero values in virtue of upper bounding the average gate fidelity between $U$ and $V$. Hence a small value of $C$ guarantees that the diagonalization $W D W^\dagger$ is an accurate approximation of the target unitary $U$. We note that this cost, Eq.~\eqref{defC} involves only local measures and hence mitigates trainability issues associated with barren plateaus \cite{mcclean2018barren,cerezo2021cost,uvarov2020barren,wang2020noise,cerezo2020impact,pesah2020absence,holmes2020barren,arrasmith2020effect,marrero2020entanglement,patti2020entanglement,holmes2021connecting,grant2019initialization}. For more details on the cost see Ref.~\cite{khatri2019quantum}. 


\medskip

\paragraph*{Training.}

The parameters $(\theta_1 , \cdots,  \theta_{18}, \gamma_1, \cdots, \gamma_3)$ are initialized to minimize the cost (\ref{defC}) with the exchange energy $J$ in (\ref{defIsing}) set to zero, which can be simulated efficiently and solved exactly.

The circuits are trained to minimise the cost using gradient descent. At each step of the training, we measure the cost and gradients $\frac{\partial C}{\partial \theta_k}$ and $\frac{\partial C}{\partial \gamma_l}$ at a particular point $(\vec{\theta}, \vec{\gamma})$ in the parameter space, and use this
to update $(\vec{\theta}, \vec{\gamma})$ according to \begin{eqnarray}
\theta_k &\leftarrow&  \theta_k - \eta \frac{\partial C}{\partial \theta_k} \\
\gamma_l &\leftarrow& \gamma_l - \eta \frac{\partial C}{\partial \gamma_l},
\end{eqnarray}
where $\eta$ is the learning rate. 
The gradients $\frac{\partial C}{\partial \theta_k}$ and $\frac{\partial C}{\partial \gamma_l}$ are measured using the parameter shift rule derived in Ref.~\cite{cirstoiu2020variational}. In each optimization step the 21 components of the gradient are measured, requiring the measurement of 156 distinct circuits. For each cost evaluation we took 8000 measurement shots. 

The learning rate was decreased as the optimization progressed according to a schedule 
\begin{equation}
\eta(j) = \frac{\eta_0}{ [1 + (j/\delta)]^{\kappa}} ,
\end{equation}
where $j \in \{0,1,2, \cdots , 16\}$ is the optimization step number. The hyperparameters  $\eta_0 \! = \!1.1$, $\kappa \! = \! 0.5,$ and $\delta \! = \! 12$ were optimized by classical simulation. Additional details about the training are provided in the Supplementary Information.

\section{Results}

\begin{figure}
\includegraphics[width=8.0cm]{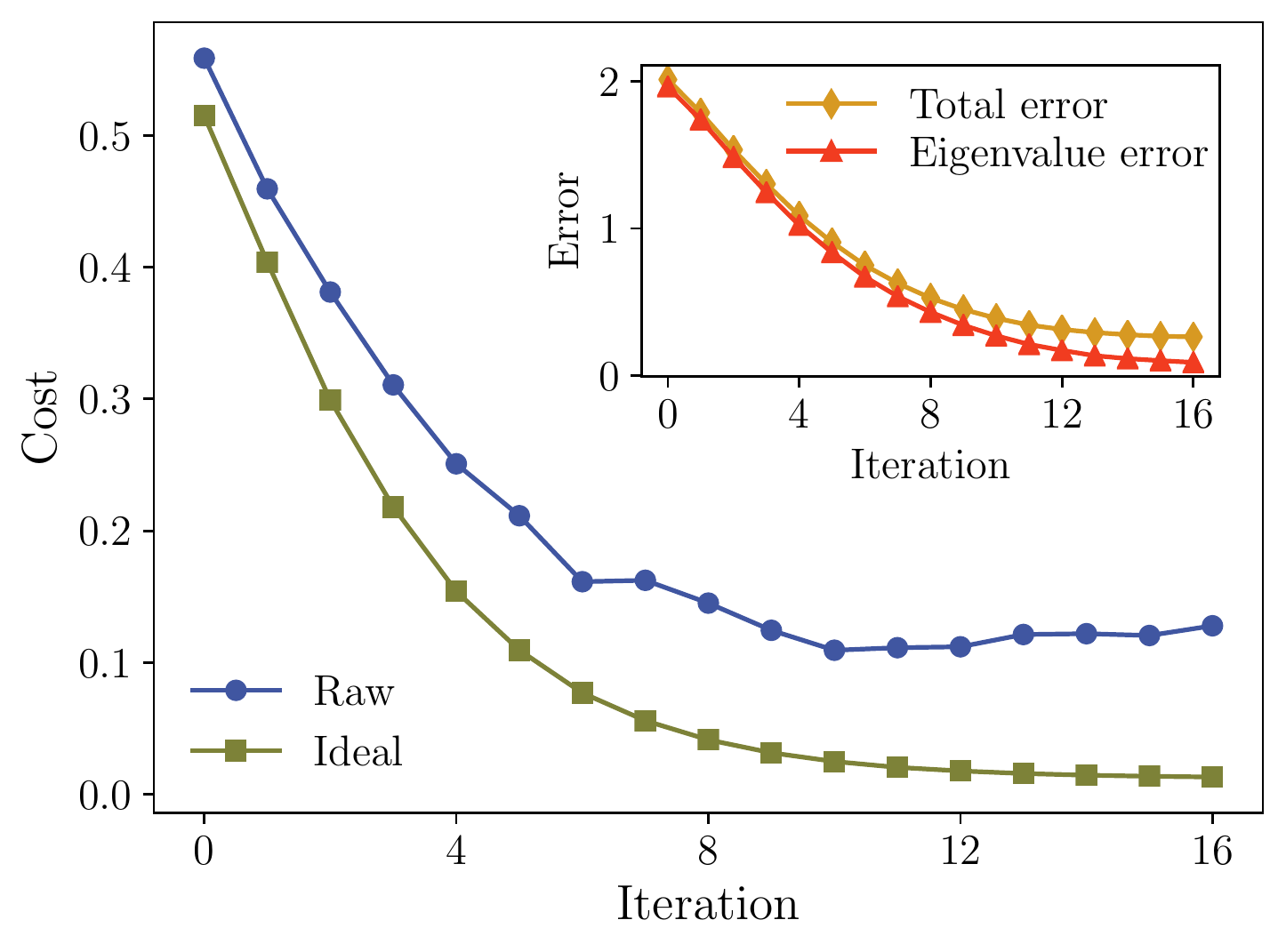} 
\caption{\textbf{Training curves.} In the main plot we show the cost as a function of the iteration step during training. The blue curve is the measured cost function $C$ and the green curve is its noise-free `ideal' value. In the inset, the yellow curve indicates the total error in the learnt unitary defined as Frobenius distances between $U$ and $V$ minimized over a global phase $e^{i\varphi}$, i.e., the quantity $\text{min}_{\varphi, \chi} || U - e^{i \varphi} V||$.  The red curve indicates the eigenvalue error, defined as the Frobenius distance between the learnt and exact diagonal also minimized over a permutation $\chi$, i.e., $\text{min}_{\varphi, \chi} || D_{\rm exact} - e^{i \varphi} \chi D \chi^\dagger||$.}
\label{training curve cost figure}
\end{figure}

To assess the quality of the optimization, the parameters found at each step of the training were used to evaluate $C$ both on the quantum computer and classically. The results are shown in Fig.~\ref{training curve cost figure}. The classical cost, which we call the noise-free or {\it ideal} cost, reflects the true accuracy of the optimized circuits. We successfully reduced the raw cost to 0.1099 corresponding to an ideal cost of 0.013. The raw cost from the quantum computer is higher than the ideal cost because of gate errors.

The inset of Fig.~\ref{training curve cost figure} confirms that the errors in $D$ and $WDW^\dagger$ are both iteratively reduced as the cost is trained. Here the Frobenius distance between $U$ and $V$ is plotted, minimized over the arbitrary global phase $e^{i\varphi}$. The ideal and learnt diagonals are also compared, accounting for the global phase $e^{i \varphi}$ and for a possible reordering, specified by a permutation matrix $\chi$. Specifically, we plot the Frobenius distance $\text{min}_{\varphi, \chi} || D_{\rm exact} - e^{i \varphi} \chi D \chi^\dagger||$ where $D_{\rm exact}$ is a diagonal matrix with $\{ \lambda^{\rm exact}_i \}$, the ordered exact spectrum of $U$, along the diagonal. This is equivalent to the sum of the eigenvalue errors $\sum_i |\lambda^{\rm exact}_i - \lambda_i e^{i \varphi_{\rm opt}}|^2$, where $\{ \lambda_i \}$ is the ordered learnt spectrum and $\varphi_{\rm opt}$ accounts for the global phase shift.

It is also interesting to monitor the training by using the measured gradient ${\vec g}_{\rm meas}$ at each step to calculate the angle between ${\vec g}_{\rm meas}$ and the exact gradient simulated classically. This is plotted in the Supplementary Information. The data confirms that the optimization is correctly moving downhill in the cost function landscape. 

\medskip

\begin{figure}
\includegraphics[width=8.0cm]{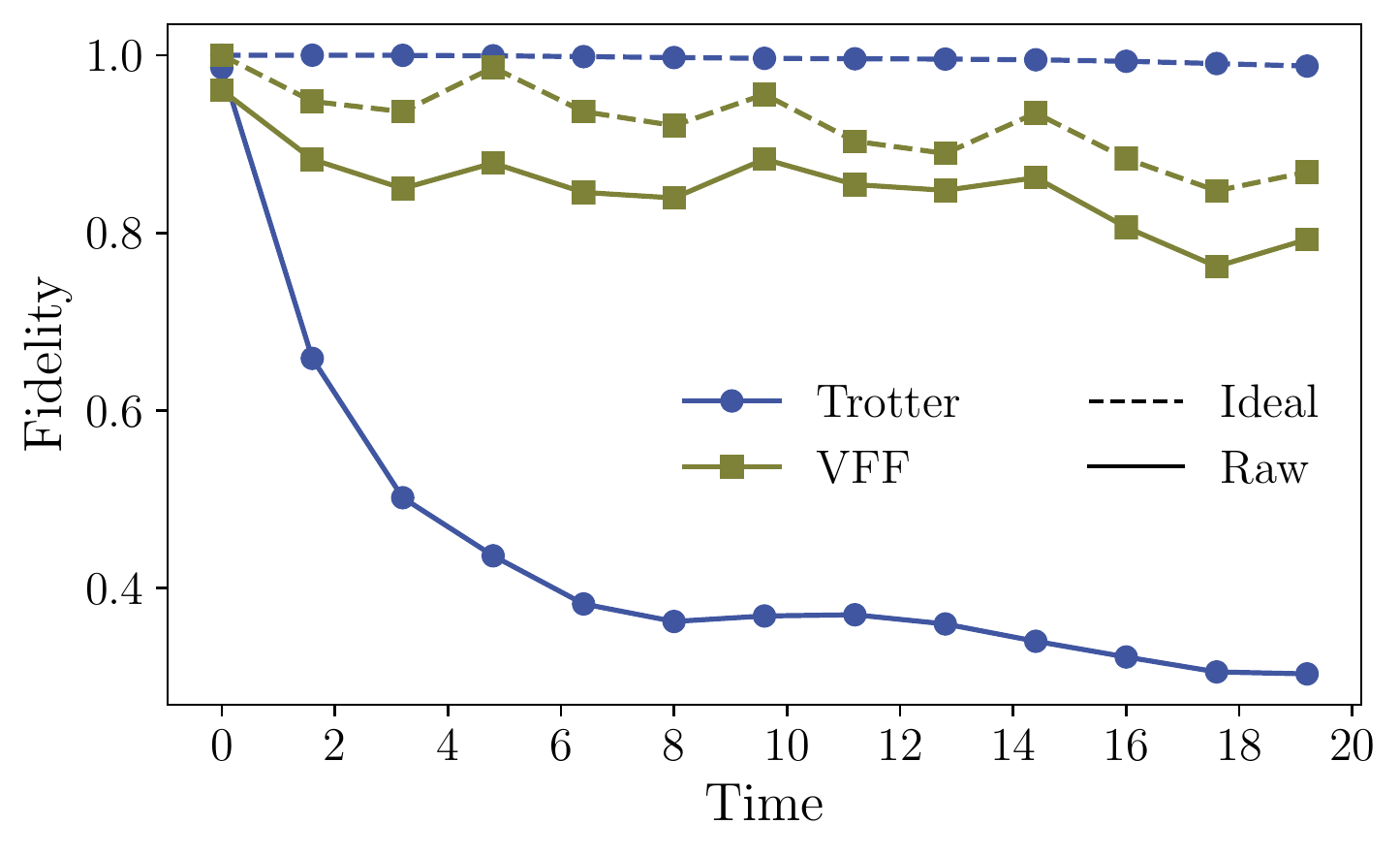} 
\caption{\textbf{Demonstration of variational fast-forwarding.} The plot shows the measured state fidelity versus evolution time. The data (circles) are obtained at multiples of $8 \, \Delta t$, out to $t \! = \! 19.2$.}
\label{fig:statefidelity}
\end{figure}

Having learned the spectral decomposition, we use the result to fast-forward the Hamiltonian simulation of the Ising model (\ref{defIsing}) for 2 spins on ibmq\_bogota. In this experiment we prepare an initial state $|+\rangle^{\otimes 2}$ and propagate it forward in time by both Trotterization and variational fast-forwarding (with backend circuit optimization disabled in both cases). The Trotterized evolution at times $t=K \Delta t$, is obtained by applying $K = 0, 8, 16, \cdots, 96$ consecutive Trotter steps from Fig.~\ref{U circuit figure}. After each step we experimentally measure the fidelity of the Trotterized evolution with the perfect evolution $e^{-iHt} |+\rangle^{\otimes 2},$ which contains no Trotter error. The variational fast-forwarding evolution at time $t$ is obtained by applying the optimized variational circuits for $W^\dagger$, $D^*$, and $W$ to $|+\rangle^{\otimes 2}$, but with $D'$s parameters $(\gamma_1, \gamma_2, \gamma_3)$ changed to 
\begin{equation}
(\gamma_1, \gamma_2, \gamma_3) \times \frac{t}{\Delta t}.
\end{equation}
The state fidelity with perfect evolution $e^{-iHt} |+\rangle^{\otimes 2}$ is also measured in this case. The results for this experimental fast-forwarding and experimental Trotter simulation are indicated by the green and blue solid lines respectively in Fig.~\ref{fig:statefidelity}. 

We compare the experimental fast forwarding to the ideal classical fast-forwarding by also measuring the noise-free fidelities obtained by classical simulation. In this ideal simulation, the initial state $|+\rangle^{\otimes 2}$ is prepared perfectly and the Trotterized evolution includes Trotter errors but no gate errors. The measurement is also assumed to be ideal. The fidelities of these ideal simulations are indicated by the dashed lines in  Fig.~\ref{fig:statefidelity}. 

The ideal Trotterized evolution is nearly perfect in Fig.~\ref{fig:statefidelity} due to the small value of $\Delta t$ and use of a second-order Trotter step. The ideal variational fast-forwarding evolution is less accurate, due to the imperfect learning of $U$'s spectral decomposition. However, the real data taken on ibmq\_bogota exhibits the opposite behavior. Whereas the variational fast-forwarding evolution is only slightly degraded by gate 
errors---because the circuit depth is independent of $t$---the Trotterized evolution quickly loses fidelity. Namely, while the fidelity of the variationally fast forwarded simulation remains above 0.7 for all 20 times steps, the fidelity of the Trotterized simulation is less than 0.4 after only 6 times steps. Thus Figure~\ref{fig:statefidelity} demonstrates a striking example of fast-forwarding as well as provides further evidence that the spectral decomposition of $U$ has been well learnt.

\medskip

In conclusion, we have experimentally demonstrated the entanglement-enhanced quantum learning of a $4 \times 4$ unitary using a diagonal ansatz containing a total of 6 CNOTs and 21 variational parameters.  A single input-output pair was used for training compared to the $2^2 = 4$ input-output pairs necessary for training without entangling ancillas. We both learn the unitary and its spectral decomposition, enabling the fast-forwarding of an Ising model. This four-qubit experiment took 8000 shots for each of the 156 independent circuit evaluations, at each of the 16 steps of the optimization algorithm, a total of $20 \times 10^7$ circuit evaluations. Thus, this experiment is among the most complex quantum machine learning demonstrations to date \cite{riste2017demonstration,schuld2017implementing,yao2017quantum,huang2018demonstration,Zhueaaw9918,kathuria2020implementation,tomesh2020coreset} and constitutes an important primitive in experimental quantum information science.

\medskip
\medskip

{\it Acknowledgments.} 
PJC and AS acknowledge initial support and MRG acknowledges support from LANL's Laboratory Directed Research and Development (LDRD) program under project number 20190065DR. Additionally, PJC and AS acknowledge that (subsequent to the above acknowledged funding) this material is based upon work supported by the U.S. Department of Energy, Office of Science, National Quantum Information Science Research Centers. ZH acknowledges support from the LANL ASC Beyond Moore's Law project. We would also like to thank Andrew Arrasmith, Lukasz Cincio, and Joe Gibbs for useful discussions.


\bibliography{ExperimentalVFF.bib}

\clearpage
\onecolumngrid


\setcounter{equation}{0}
\setcounter{figure}{0}
\setcounter{table}{0}
\setcounter{page}{1}
\setcounter{section}{0}
\setcounter{secnumdepth}{4}

\makeatletter
\renewcommand{\thesection}{\arabic{section}}
\renewcommand{\theequation}{S\arabic{equation}}
\renewcommand{\thefigure}{S\arabic{figure}}
\renewcommand{\bibnumfmt}[1]{[S#1]}

\begin{center}
\Large{ Supplementary Information for \\ ``Experimental Quantum Learning of a Spectral Decomposition''}
\end{center}

\vspace{1cm}
\twocolumngrid

This document provides additional details about the experimental results.
In Sec.~\ref{Qubits} we describe the online superconducting qubits used in the experiment and give calibration results (gate errors, coherence times, and single-qubit measurement  errors) provided by the backend. In Sec.~\ref{training section} we provide additional details about the circuit training.

\section{Qubits}\label{Qubits}

\begin{figure}[b]
\includegraphics[width=8.0cm]{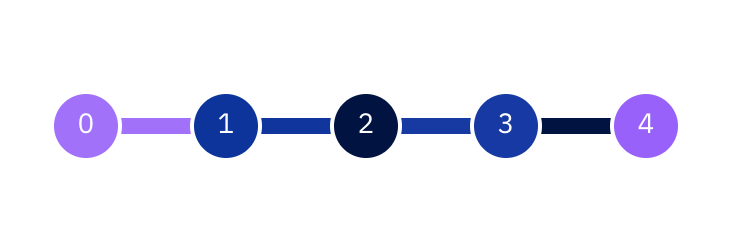} 
\caption{Layout of IBM Q device ibmq\_bogota. In this work we use qubits $Q_0$, $Q_1$, $Q_2$ and $Q_3$ for the training and qubits $Q_1$ and $Q_2$ for the Trotter and VFF comparison.}
\label{bogota chain figure}
\end{figure}

Data was taken on the IBM Q processor ibmq\_santiago using the {BQP} software package developed by Geller and colleagues. BQP is a Python package developed to design, run, and analyze complex quantum computing and quantum information experiments using commercial backends. We demonstrate the learning of a spectral decomposition using the qubits shown in Fig.~\ref{bogota chain figure}. Calibration data supplied by the backend is summarized in Table \ref{calibrationDataTable}. Here $T_{1,2}$ are the standard Markovian decoherence times, and
\begin{equation}
\epsilon = \frac{T(0|1) + T(1|0)}{2}
\end{equation}
is the single-qubit state-preparation and measurement (SPAM) error, averaged over initial states. The $U_2$ error column gives the single-qubit gate error measured by randomized benchmarking. The CNOT errors are also measured by randomized benchmarking.

\begin{table}[htb]
\centering
\caption{Calibration data provided by IBM Q for the  ibmq\_bogota chip during the period of data acquisition.}
\begin{tabular}{c}
\begin{tabular}{|c|c|c|c|c|}
\hline
Qubit & $T_1 \ (\mu s)$ & $T_2  \ (\mu s)$ & SPAM error $\epsilon$ &  $U_2$ error \\
\hline 
$Q_0$ & 136.6  & 178.0 & 0.056 & 3.70e-4  \\
$Q_1$ & 132.3  & 117.4 & 0.045 & 2.01e-4  \\
$Q_2$ & 65.4 & 132.4& 0.024 & 2.48e-4  \\
$Q_3$ & 103.6  & 158.3 & 0.020 & 4.74e-4  \\
\hline 
\end{tabular}
\\
\begin{tabular}{|c|c|}
\hline 
CNOT gates &  CNOT error  \\
\hline 
\begin{tabular}{c|c} ${\rm CNOT}_{0,1}$ &  ${\rm CNOT}_{1,0}$  \\ \end{tabular} &   1.04e-3   \\
\hline 
\begin{tabular}{c|c} ${\rm CNOT}_{1,2}$ &  ${\rm CNOT}_{2,1}$  \\ \end{tabular} &   8.12e-3    \\
\hline 
\begin{tabular}{c|c} ${\rm CNOT}_{2,3}$ &  ${\rm CNOT}_{3,2}$  \\ \end{tabular} &   3.81e-2    \\
\hline 
\end{tabular} 
\\
\end{tabular}
\label{calibrationDataTable}
\end{table}

\section{Training}
\label{training section}

\begin{figure*}
\includegraphics[width=10.0cm]{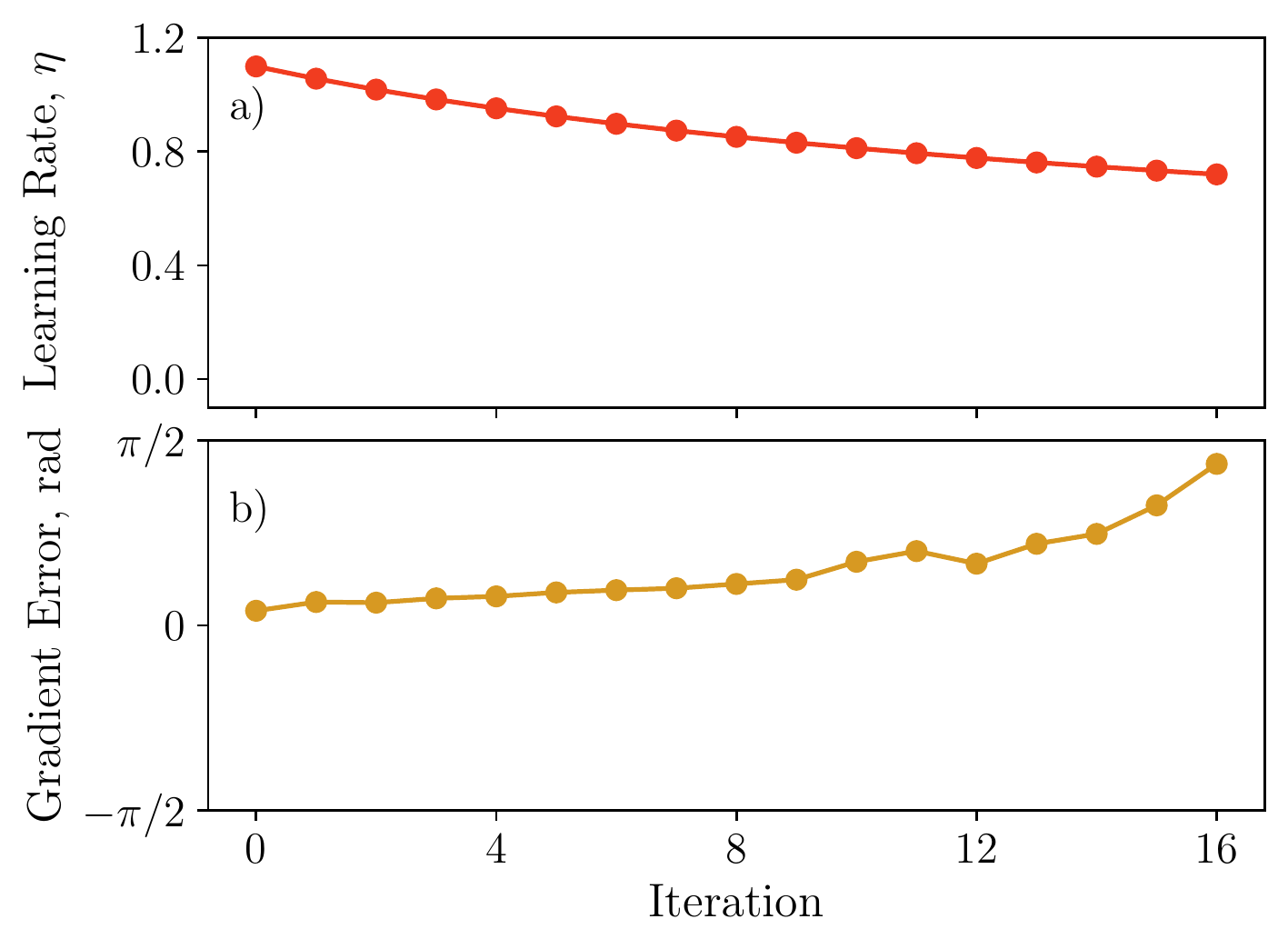} 
\caption{(a) Learning rate versus optimization step during training by gradient descent. This rate obeys $\eta(j) = \frac{\eta_0}{ [1 + (j/\delta)]^{\kappa}}$
where $j \in \{0,1,2, \cdots , 16\}$ is the optimization step number and the hyperparameters  $\eta_0 \! = \!1.1$, $\kappa \! = \! 0.5,$ and $\delta \! = \! 12$ were optimized by classical simulation.
(b) Error in the measured gradient direction during training. Here we plot the angle between gradient measured on ibmq\_bogota during training, and true gradient calculating classically for the same set of parameters, for each optimization step.}
\label{learning rate figure}
\end{figure*}

The variational circuits for $W$ and $D$ were trained by gradient descent
\begin{equation}
\begin{aligned}
&\theta_k \leftarrow \theta_k - \eta \frac{\partial C_{\rm LHST}}{\partial \theta_k} \\
&\gamma_l \leftarrow \gamma_l - \eta \frac{\partial C_{\rm LHST}}{\partial \gamma_l},
\end{aligned}
\label{GD update rule}
\end{equation}
where $\theta_k$ and $\gamma_l$ are the variational parameters for the $W$ and $D$ circuits, respectively, and $\eta$ is the learning rate.  We used the 
variable learning rate plotted in Fig.~\ref{learning rate figure}a), which was optimized by classical simulation. At each step of the training, we measure the 
cost $C_{\rm LHST}$ and gradients $\partial C_{\rm LHST} / \partial \theta_k$ and $\partial C_{\rm LHST} / \partial \gamma_l$ at a particular point $(\theta_1 , \cdots,  \theta_{18}, \gamma_1, \cdots, \gamma_3)$ in parameter space, and use this to update $(\theta_1 , \cdots,  \theta_{18}, \gamma_1, \cdots, \gamma_3)$ according to (\ref{GD update rule}). We also calculate the angle between the measured gradient and the exact gradient simulated classically, which is plotted in Fig.~\ref{learning rate figure}b). This data shows that the measured gradient is correctly pointing uphill until the end of the training when a local minimum is reached and the gradient becomes small and noisy.

\end{document}